%#!tex D5-ALE.tex
%% Last Modified: Wed Oct 26 08:57:36 2005.

%%%%%%%%%%%%%%%%%%%%%%%%%%%%%%%%%%%%%%%%%%%%%%%%%%%%%%%%%%%%%%%%%
%                                                               %
%                          NS5                                  %
%                                                               %
%                  Kazumi Okuyama (UBC)                         %
%                                                               %
%%%%%%%%%%%%%%%%%%%%%%%%%%%%%%%%%%%%%%%%%%%%%%%%%%%%%%%%%%%%%%%%%

%\input harvmac

%\def\mydraft{label}

\input lanlmac
\input amssym
\input epsf

%Macro for figure
\newcount\figno
\figno=0
\def\fig#1#2#3{
\par\begingroup\parindent=0pt\leftskip=1cm\rightskip=1cm\parindent=0pt
\baselineskip=13pt
\global\advance\figno by 1
\midinsert
\epsfxsize=#3
\centerline{\epsfbox{#2}}
\vskip 12pt
%\centerline{{\bf Fig. \the\figno:~~} #1}\par
{\bf Fig. \the\figno:~~} #1 \par
\endinsert\endgroup\par
}
\def\figlabel#1{\xdef#1{\the\figno}}

%Macro

\def\th{\theta}

\def\ep{\epsilon}

\def\S{{\bf S}}

\def\tr{{\rm tr}}
\def\Tr{{\rm Tr}}
\def\hf{{1\over 2}}

\def\o{\over}
\def\Up{\Upsilon}
\def\til#1{\widetilde{#1}}
\def\si{\sigma}
\def\Si{\Sigma}
\def\b#1{\overline{#1}}
\def\del{\partial}

\def\lf{\left}
\def\ri{\right}
\def\riya{\rightarrow}

\def\lrya{\leftrightarrow}

\def\La{\Lambda}
\def\h#1{\widehat{#1}}

\def\Ga{\Gamma}

\def\om{\omega}

\def\Om{\Omega}
\def\dag{\dagger}
\def\rt#1{\sqrt{#1}}

\def\sitarel#1#2{\mathrel{\mathop{\kern0pt #1}\limits_{#2}}}

\def\cob{\delta}

% \newsec{References}
\lref\DouglasUZ{
  M.~R.~Douglas,
  ``Gauge Fields and D-branes,''
  J.\ Geom.\ Phys.\  {\bf 28}, 255 (1998)
  [arXiv:hep-th/9604198].
  %%CITATION = HEP-TH 9604198;%%
}
\lref\WittenTZ{
  E.~Witten,
  ``Sigma models and the ADHM construction of instantons,''
  J.\ Geom.\ Phys.\  {\bf 15}, 215 (1995)
  [arXiv:hep-th/9410052].
  %%CITATION = HEP-TH 9410052;%%
}
\lref\WittenYU{
  E.~Witten,
  ``On the conformal field theory of the Higgs branch,''
  JHEP {\bf 9707}, 003 (1997)
  [arXiv:hep-th/9707093].
  %%CITATION = HEP-TH 9707093;%%
}
\lref\DiaconescuGU{
  D.~E.~Diaconescu and N.~Seiberg,
  ``The Coulomb branch of (4,4) supersymmetric field theories in two
  dimensions,''
  JHEP {\bf 9707}, 001 (1997)
  [arXiv:hep-th/9707158].
  %%CITATION = HEP-TH 9707158;%%
}
\lref\DouglasVU{
  M.~R.~Douglas, J.~Polchinski and A.~Strominger,
  ``Probing five-dimensional black holes with D-branes,''
  JHEP {\bf 9712}, 003 (1997)
  [arXiv:hep-th/9703031].
  %%CITATION = HEP-TH 9703031;%%
}
\lref\DouglasSW{
  M.~R.~Douglas and G.~W.~Moore,
  ``D-branes, Quivers, and ALE Instantons,''
  arXiv:hep-th/9603167.
  %%CITATION = HEP-TH 9603167;%%
}
\lref\JohnsonPY{
  C.~V.~Johnson and R.~C.~Myers,
  ``Aspects of type IIB theory on ALE spaces,''
  Phys.\ Rev.\ D {\bf 55}, 6382 (1997)
  [arXiv:hep-th/9610140].
  %%CITATION = HEP-TH 9610140;%%
}
\lref\TongRQ{
  D.~Tong,
  ``NS5-branes, T-duality and worldsheet instantons,''
  JHEP {\bf 0207}, 013 (2002)
  [arXiv:hep-th/0204186].
  %%CITATION = HEP-TH 0204186;%%
}
\lref\HarveyAB{
  J.~A.~Harvey and S.~Jensen,
  ``Worldsheet instanton corrections to the Kaluza-Klein monopole,''
  arXiv:hep-th/0507204.
  %%CITATION = HEP-TH 0507204;%%
}
\lref\OkuyamaGX{
  K.~Okuyama,
  ``Linear sigma models of H and KK monopoles,''
  JHEP {\bf 0508}, 089 (2005)
  [arXiv:hep-th/0508097].
  %%CITATION = HEP-TH 0508097;%%
}
\lref\AharonyDW{
  O.~Aharony and M.~Berkooz,
  ``IR dynamics of d = 2, N = (4,4) gauge theories and DLCQ of 'little  string
  theories',''
  JHEP {\bf 9910}, 030 (1999)
  [arXiv:hep-th/9909101].
  %%CITATION = HEP-TH 9909101;%%
}
\lref\WittenYC{
  E.~Witten,
  ``Phases of N = 2 theories in two dimensions,''
  Nucl.\ Phys.\ B {\bf 403}, 159 (1993)
  [arXiv:hep-th/9301042].
  %%CITATION = HEP-TH 9301042;%%
}
\lref\LarsenDH{
  F.~Larsen and E.~J.~Martinec,
  ``Currents and moduli in the (4,0) theory,''
  JHEP {\bf 9911}, 002 (1999)
  [arXiv:hep-th/9909088].
  %%CITATION = HEP-TH 9909088;%%
}
\lref\KutasovZH{
  D.~Kutasov, F.~Larsen and R.~G.~Leigh,
  ``String theory in magnetic monopole backgrounds,''
  Nucl.\ Phys.\ B {\bf 550}, 183 (1999)
  [arXiv:hep-th/9812027].
  %%CITATION = HEP-TH 9812027;%%
}
\lref\SugawaraQP{
  Y.~Sugawara,
  ``N = (0,4) quiver SCFT(2) and supergravity on AdS(3) x S(2),''
  JHEP {\bf 9906}, 035 (1999)
  [arXiv:hep-th/9903120].
  %%CITATION = HEP-TH 9903120;%%
}
\lref\BenaAY{
  I.~Bena and P.~Kraus,
  ``Microstates of the D1-D5-KK system,''
  Phys.\ Rev.\ D {\bf 72}, 025007 (2005)
  [arXiv:hep-th/0503053].
  %%CITATION = HEP-TH 0503053;%%
}
\lref\SeibergXZ{
  N.~Seiberg and E.~Witten,
  ``The D1/D5 system and singular CFT,''
  JHEP {\bf 9904}, 017 (1999)
  [arXiv:hep-th/9903224].
  %%CITATION = HEP-TH 9903224;%%
}
\lref\LarsenUK{
  F.~Larsen and E.~J.~Martinec,
  ``U(1) charges and moduli in the D1-D5 system,''
  JHEP {\bf 9906}, 019 (1999)
  [arXiv:hep-th/9905064].
  %%CITATION = HEP-TH 9905064;%%
}
\lref\MaldacenaDE{
  J.~M.~Maldacena, A.~Strominger and E.~Witten,
  ``Black hole entropy in M-theory,''
  JHEP {\bf 9712}, 002 (1997)
  [arXiv:hep-th/9711053].
  %%CITATION = HEP-TH 9711053;%%
}
\lref\MinasianQN{
  R.~Minasian, G.~W.~Moore and D.~Tsimpis,
  ``Calabi-Yau black holes and (0,4) sigma models,''
  Commun.\ Math.\ Phys.\  {\bf 209}, 325 (2000)
  [arXiv:hep-th/9904217].
  %%CITATION = HEP-TH 9904217;%%
}
\lref\DistlerMI{
  J.~Distler,
  ``Notes on (0,2) superconformal field theories,''
  arXiv:hep-th/9502012.
  %%CITATION = HEP-TH 9502012;%%
}
\lref\AdamsZY{
  A.~Adams, A.~Basu and S.~Sethi,
  ``(0,2) duality,''
  Adv.\ Theor.\ Math.\ Phys.\  {\bf 7}, 865 (2004)
  [arXiv:hep-th/0309226].
  %%CITATION = HEP-TH 0309226;%%
}
\lref\NekrasovSS{
  N.~Nekrasov and A.~Schwarz,
  ``Instantons on noncommutative R**4 and (2,0) superconformal six  dimensional
  theory,''
  Commun.\ Math.\ Phys.\  {\bf 198}, 689 (1998)
  [arXiv:hep-th/9802068].
  %%CITATION = HEP-TH 9802068;%%
}
\lref\CallanAT{
  C.~G.~.~Callan, J.~A.~Harvey and A.~Strominger,
  ``Supersymmetric string solitons,''
  arXiv:hep-th/9112030.
  %%CITATION = HEP-TH 9112030;%%
}
\lref\JohnsonGA{
  C.~V.~Johnson, R.~R.~Khuri and R.~C.~Myers,
  ``Entropy of 4D Extremal Black Holes,''
  Phys.\ Lett.\ B {\bf 378}, 78 (1996)
  [arXiv:hep-th/9603061].
  %%CITATION = HEP-TH 9603061;%%
}
\lref\GukovYM{
  S.~Gukov, E.~Martinec, G.~W.~Moore and A.~Strominger,
  ``The search for a holographic dual to AdS(3) x S**3 x S**3 x S**1,''
  arXiv:hep-th/0403090.
  %%CITATION = HEP-TH 0403090;%%
}
\lref\GibbonsNT{
  G.~W.~Gibbons and P.~Rychenkova,
  ``HyperKaehler quotient construction of BPS monopole moduli spaces,''
  Commun.\ Math.\ Phys.\  {\bf 186}, 585 (1997)
  [arXiv:hep-th/9608085].
  %%CITATION = HEP-TH 9608085;%%
}
\lref\SaxenaUK{
  A.~Saxena, G.~Potvin, S.~Giusto and A.~W.~Peet,
  ``Smooth geometries with four charges in four dimensions,''
  arXiv:hep-th/0509214.
  %%CITATION = HEP-TH 0509214;%%
}

%%%%%%%%%%%%%%%%%%%%%%%%%%%%%%%%%%%%%%%%%%%%%%%%%%%%%%%%%%%%%%%%%
%                      Title Page                               %
%%%%%%%%%%%%%%%%%%%%%%%%%%%%%%%%%%%%%%%%%%%%%%%%%%%%%%%%%%%%%%%%%
\Title{             
                                             \vbox{
                                             \hbox{hep-th/0510195}}}
{\vbox{
\centerline{D1-D5 on ALE Space}
}}

\vskip .2in

\centerline{Kazumi Okuyama}

\vskip .2in

%\vskip 2cm
\centerline{Department of Physics and Astronomy, 
University of British Columbia} 
\centerline{Vancouver, BC, V6T 1Z1, Canada}
\centerline{\tt kazumi@phas.ubc.ca}
\vskip 3cm
\noindent

%%abstract
We construct a two-dimensional ${\cal N}=(0,4)$ quiver gauge theory
on D1-brane probing D5-branes on ALE space, 
and study its IR behavior.
This can be thought of as a 
gauged linear sigma model for the NS5-branes on ALE space.

\Date{October 2005}

\vfill
\vfill

\newsec{Introduction}
The D1-D5-KK system (KK = Kaluza-Klein monopole) is a
1/8 BPS
configuration in type IIB string theory.
We are interested in the two-dimensional ${\cal N}=(0,4)$
gauge theory on the D1-brane of this system.
In the limit where the KK-monopole is replaced by the 
orbifold ${\Bbb C}^2/{\Bbb Z}_n$,
it is straightforward to construct the theory on D1-brane following the 
standard procedure of Douglas and Moore \DouglasSW. 
Namely, we consider the ${\Bbb Z}_n$ orbifolding of
the ${\cal N}=(4,4)$ $U(Q_1)$ gauge theory with one adjoint and $Q_5$
fundamental hypermultiplets
coming from the D1-D5 system.
The resulting theory is an ${\cal N}=(0,4)$ quiver gauge theory.
As compared to the study of gauge theory of the D1-D5 system 
\refs{\WittenYU\DiaconescuGU\SeibergXZ\LarsenUK{--}\AharonyDW},
the ${\cal N}=(0,4)$ gauge theory of the D1-D5-KK system is less understood.
See \refs{\KutasovZH\SugawaraQP\LarsenDH\BenaAY{--}\SaxenaUK} 
for some of the
works related to this system.

The D1-D5-KK system is related to the various brane configurations
by duality. Obviously it is S-dual to the F1-NS5-KK system, 
and it is also dual to the triple intersection of M5-branes 
\refs{\MaldacenaDE,\MinasianQN}.
In general, ${\cal N}=(0,4)$ CFTs appear in many places in
string theory; some of these are related to the D1-D5-KK system 
and some are not. 
%For example, the D1-D5-KK system can be transformed to
%the D1-D5-P system by a chain of duality.
The examples of $(0,4)$ preserving configuration that
are not directly related to the D1-D5-KK system include
the D1-D5-D9 system described 
by the $(0,4)$ ADHM sigma model \refs{\DouglasUZ,\WittenTZ}, and
the intersecting brane configuration of 
D1-D5-D5 \GukovYM.
 
One of the motivation to study the ${\cal N}=(0,4)$ quiver gauge 
theory
is that it can be thought of as a gauged linear sigma model (GLSM) 
for the
NS5-branes on ALE space.
In general, GLSM  is a quite useful tool
to understand the moduli space of CFTs and the relations among them 
\WittenYC.
In \refs{\TongRQ,\HarveyAB,\OkuyamaGX}, 
NS5-branes on $\S^1$ and their relation to the T-dual
KK-monopoles are studied by using an ${\cal N}=(4,4)$ GLSM.
It is very interesting to find GLSMs describing
NS5-branes in other backgrounds.
Our ${\cal N}=(0,4)$ quiver gauge theory is 
such an example.

This paper is organized as follows.
In section 2, we construct the ${\cal N}=(0,4)$ quiver gauge theory
describing the D1-D5 branes on ${\Bbb C}^2/{\Bbb Z}_n$.
In section 3, we study the Higgs branch of this theory.
In section 4, we compute the 1-loop correction to the Coulomb branch
metric. Section 5 is discussions.

\newsec{D1-D5 on ${\Bbb C}^2/{\Bbb Z}_n$}
In this section, we construct the Lagrangian on
the D1-brane probe for the D5-branes on ${\Bbb C}^2/{\Bbb Z}_n$. 
\subsec{Symmetries}
To find the Lagrangian on the D1-brane, we first summarize the symmetries
of the system.
Let us consider a configuration of D1-branes and D5-branes extending in
the $x^0x^1$ and $x^0x^1\cdots x^5$ directions, respectively.
%This configuration preserves 1/4 of spacetime SUSY.
We can perform the orbifolding of the transverse directions 
${\Bbb C}^2(=x^6x^7x^8x^9)$ by ${\Bbb Z}_n$ such that the resulting 
configuration is 1/8 BPS.
This can be seen by writing the BPS condition for the
supersymmetry generator $\ep_LQ_L+\ep_RQ_R$
\eqn\epLR{\eqalign{
{\rm D1:}~&\ep_L=\Ga^0\Ga^1\ep_R \cr
{\rm D5:}~&\ep_L=\Ga^0\Ga^1\Ga^2\Ga^3\Ga^4\Ga^5\ep_R \cr
{\rm KK:}~&\ep_L=\Ga^0\Ga^1\Ga^2\Ga^3\Ga^4\Ga^5\ep_L,\quad
\ep_R=\Ga^0\Ga^1\Ga^2\Ga^3\Ga^4\Ga^5\ep_R
}}
where $\ep_L,\ep_R$ are ten-dimensional Majorana-Weyl spinors with the same
chirality.
From these relations, it follows that 
the unbroken supersymmetry in 
the $x^0x^1$ space is 
chiral $\ep_R=\ep_L=\Ga^0\Ga^1\ep_L$.
One can check that it is
a two-dimensional ${\cal N}=(0,4)$.

Before orbifolding, the massless open string spectrum
on the D1-brane is decomposed into various representations
under the symmetry $SO(1,1)_{01}\times SO(4)_{2345}\times SO(4)_{6789}$.
Following \DouglasUZ, we introduce the indices
$(A',\til{A}')$ and $(A,Y)$ to represent
 the fundamental ${\bf 2}$ of various $SU(2)$ groups:
\eqn\SUtwoind{\eqalign{
SO(4)_{2345}&=SU(2)_{A'}\times SU(2)_{\til{A}'},\cr
SO(4)_{6789}&=SU(2)_{A}\times SU(2)_{Y}.
}}
The massless modes on the D1-brane
coming from the 1-1 string and 1-5 string are summarized as 
\DouglasUZ\foot{We flipped the $SO(1,1)$ chirality from 
\DouglasUZ\ for later convenience.
This is merely a convention of calling the remaining supersymmetry
$(0,4)$ or $(4,0)$.}
\eqn\secone{\eqalign{
\matrix{&b^{AY}&\psi_{+}^{A'Y}\cr 
1-1:~~&b^{A'\til{A}'}&\psi_{+}^{A\til{A}'}\cr 
&A_{--},A_{++}&\psi_-^{AA'},
\psi_-^{\til{A}'Y}\cr
1-5:~~&H^{A'}&\chi_+^{A},\chi_-^{Y}}
}}
where $A_{\pm\pm}=A_0\pm A_1$.
This spectrum represent the ${\cal N}=(4,4)$ $U(Q_1)$ gauge theory
with one adjoint hypermultiplet and $Q_5$ fundamental
hypermultiplets.
$b^{AY}$ is the scalar in the ${\cal N}=(4,4)$ 
vectormultiplet representing the position of D1-brane 
in the $(6789)$ directions,
while $b^{A'\til{A}'}$ is the scalar in the adjoint hypermultiplet
corresponding to the $(2345)$ directions.
$SU(2)_{A'}$ is usually denoted $SU(2)_R$ since the scalars in the
hypermultiplets $b^{A'\til{A}'}$ and $H^{A'}$ are doublets  under 
this $SU(2)$.
The ${\cal N}=(4,4)$ supersymmetry is generated by
the supercharges
\eqn\susypm{
Q^{AA'}_+,\quad Q^{A'Y}_-.
}
For example, the super transformation of $b^{AY}$ is
\eqn\trfb{
Q^{AA'}_{+}b^{~~Y}_A=\psi_{+}^{A'Y},\quad
 Q^{A'Y}_-b^A_{~~Y}=\psi_{-}^{AA'}.
}

Now we consider the orbifolding of ${\Bbb C}^2={\Bbb R}^4_{6789}$ 
by ${\Bbb Z}_n$.
We would like to preserve the right moving $(0,4)$ supersymmetry generated
by $Q^{AA'}_+$ and break the left moving one $Q^{A'Y}_-$.
Therefore, we embed the orbifold group ${\Bbb Z}_n$ in $SU(2)_Y$:
\eqn\embedZn{
{\Bbb Z}_n\subset SU(2)_Y\subset SO(4)_{6789}.
}
Thus the nontrivial orbifold action is on the $Y$-index
\eqn\omact{
(b^{AY},\psi_+^{A'Y},\psi_-^{\til{A}'Y},\chi_-^Y)
\riya 
(\om^{Y}b^{AY},\om^{Y}\psi_+^{A'Y},\om^{Y}\psi_-^{\til{A}'Y},\om^{Y}\chi_-^Y)
}
where $\om=e^{2\pi i/n}$ and $Y=\pm$, and the rest of the fields in
\secone\ are invariant under the ${\Bbb Z}_n$.

To construct the gauge theory on the D1-brane, 
we follow the general procedure 
\DouglasSW. We extend the Chan-Paton factors ${\cal H}_{D1}$ and
${\cal H}_{D5}$ associated with the D1-branes and the D5-branes
by tensoring ${\Bbb C}^n$. Physically this corresponds to going
to the covering space of ${\Bbb C}^2/{\Bbb Z}_n$.
Then we extend the orbifold action 
\omact\ by acting ${\Bbb Z}_n$ on 
the Chan-Paton factor ${\Bbb C}^n$ as well.
We choose this action on ${\Bbb C}^n$ 
to be the regular representation of ${\Bbb Z}_n$.
We also set ${\cal H}_{D1}={\Bbb C}$ and ${\cal H}_{D5}={\Bbb C}^k$,
so that the gauge group becomes product of $U(1)$'s.

\fig{The quiver diagram for D1-D5 on ${\Bbb C}^2/{{\Bbb Z}_n}$.
The inner (outer) quiver corresponds to the D1-branes (D5-branes).
The inner quiver is the $\h{A}_{n-1}$ Dynkin diagram (only three nodes 
are drawn in this figure).
The links in the inner quiver represent a non-chiral ${\cal N}=(4,4)$
matter content, while the modes coming from the 1-5 string become chiral
after orbifold projection. The dashed lines  represent
the left moving Fermi multiplets 
$(\La_{a+1}^{Q{f_{a}}},\La_{f_{a+1},a}^{\til{Q}})$.
}
{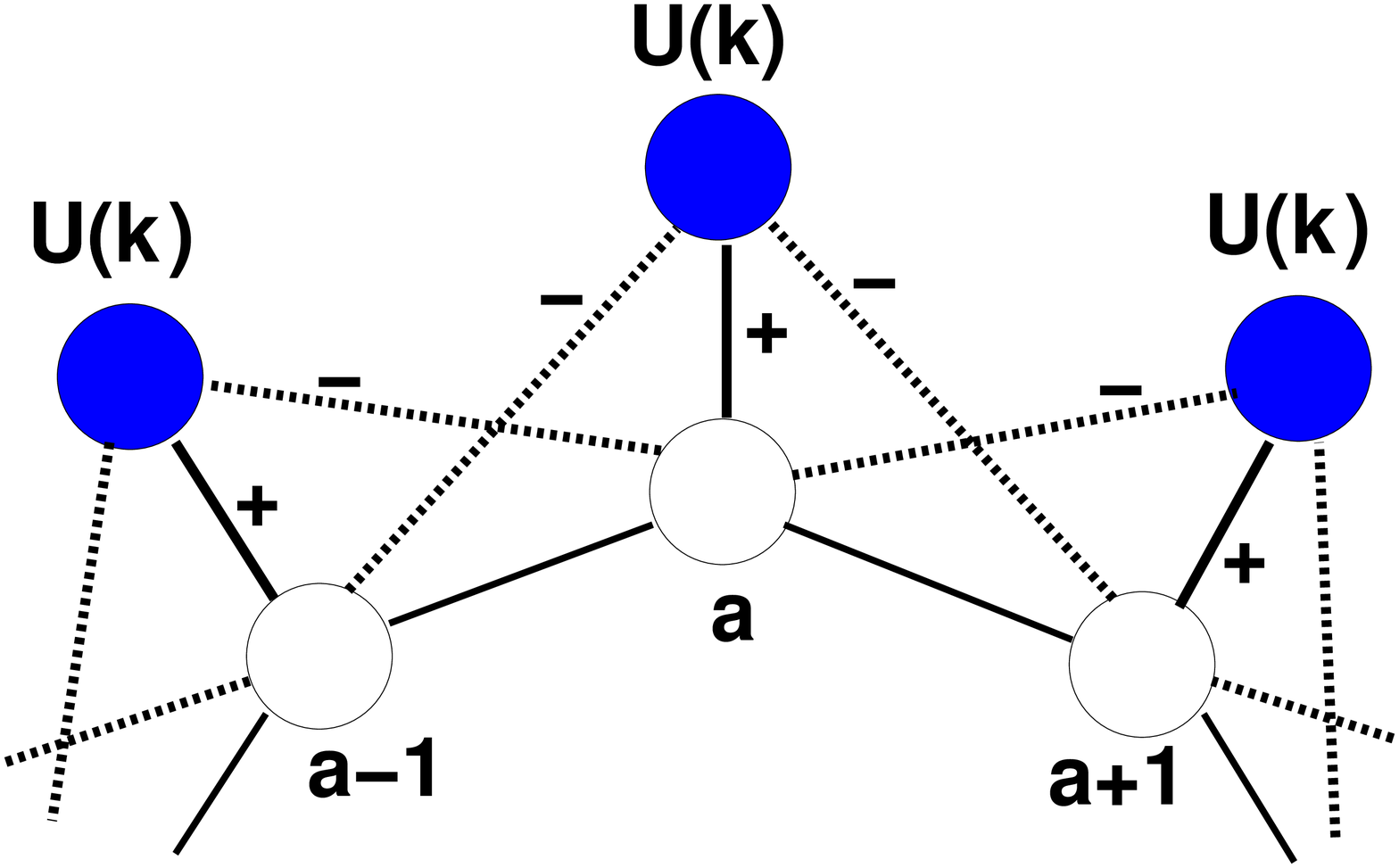}{8cm}

The resulting ${\Bbb Z}_n$ invariant spectrum are
summarized by a quiver diagram (see Fig. 1).
This is a two sets of $\h{A}_{n-1}$ Dynkin diagram
connected by links. The inner (outer) quiver corresponds
to the D1-brane (D5-brane) \foot{In Fig. 1, we didn't draw the links 
in the outer quiver, since they don't correspond to the dynamical fields on
the D1-brane gauge theory.}. The inner quiver represents
the theory on the D1-branes on ${\Bbb C}^2/{\Bbb Z}_n$
constructed in \JohnsonPY, which is non-chiral and invariant
under the ${\cal N}=(4,4)$ supersymmetry.  This theory is well-known to lead 
to the hyperK\"{a}hler quotient construction of
$A_{n-1}$ ALE space.
The left-right asymmetry comes
from the orbifold action on the 1-5 string.
As we can see, $H^{A'}$ and $\chi_+^A$ in \secone\ are 
invariant under ${\Bbb Z}_n$
while $\chi_{-}^{Y}$ transforms non-trivially \omact.
This difference leads to the peculiar structure of the links connecting
the inner and outer quivers as shown in Fig. 1.

\subsec{The Lagrangian in the ${\cal N}=(0,2)$ Language}
To write down the Lagrangian explicitly, it is useful to use
the ${\cal N}=(0,2)$ superspace language. We follow the notation
in \WittenYC \foot{See also \refs{\DistlerMI,\AdamsZY} 
for useful reviews on $(0,2)$ linear sigma models.}.

Before orbifolding, the theory representing the D1-D5 system is an 
${\cal N}=(4,4)$ $U(Q_1)$ gauge theory with one adjoint and $Q_5$
fundamental hypermultiplets.
In the language of ${\cal N}=(2,2)$ superfields,
the matter content of this theory
is the vector multiplet $(\Si,\Phi)$, the adjoint hypermultiplet $(B,\til{B})$
and the fundamental hypermultiplets $(Q,\til{Q})$. 
All superfields are 
${\cal N}=(2,2)$ chiral superfields,
except for $\Si$ which is an ${\cal N}=(2,2)$ twisted chiral superfield.
These fields correspond to the notation \secone\
in the previous section as
\eqn\corresp{
b^{AY}\lrya (\Si,\Phi),\quad
b^{A'\til{A}'}\lrya(B,\til{B}),\quad
H^{A'}\lrya (Q,\til{Q}).
}

To proceed, we need to know the orbifold action in
the ${\cal N}=(2,2)$ language.
Without loss of generality, we can choose 
the superspace coordinate $\th^-$ to carry
the index $Y=+$. Therefore, the ${\Bbb Z}_n$ action in the ${\cal N}=(2,2)$
language is given by
\eqn\Zntwotwo{\eqalign{
\Si(\th^+,\bar{\th}^-),~\Phi(\th^+,\th^-)&\riya\om\Si(\th^+,
\om^{-1}\bar{\th}^{-}),~\om^{-1}\Phi(\th^+,\om\th^-) \cr
B(\th^+,\th^-),~\til{B}(\th^+,\th^-)&\riya
B(\th^+,\om\th^-),~\til{B}(\th^+,\om\th^-) \cr
Q(\th^+,\th^-),~\til{Q}(\th^+,\th^-)&\riya
Q(\th^+,\om\th^-),~\til{Q}(\th^+,\om\th^-)
}} 
This condition can be written down in the ${\cal N}=(0,2)$ language by
decomposing the ${\cal N}=(2,2)$ superfields into the ${\cal N}=(0,2)$
superfields:
\eqn\chFermi{\eqalign{
\Si_{(2,2)}(\th^+,\bar{\th}^-)&\sim\Si-\rt{2}\bar{\th}^-\Up \cr
\Phi^i_{(2,2)}(\th^+,\bar{\th}^-)&\sim\Phi^i+\rt{2}\th^-\La^i
}}
where $\Si$ and $\Phi^i$ are the $(0,2)$ chiral superfields,
$\Up$ is the $(0,2)$ gauge superfield, and $\La^i$ is the Fermi superfield.

The resulting theory is an ${\cal N}=(0,4)$ gauge theory having
$\prod_{a=1}^nU(1)_a$ as the
gauge group and $\prod_{a=1}^nU(k)_a$ as the flavor symmetry.
The kinetic term of the orbifolded theory is given by
\eqn\Lagkin{\eqalign{
{\cal L}_{kin}=-\hf \sum_{a=1}^n\int d\th^+d\bar{\th}^+\Bigg[
-{i\o e^2}&\b{\Si}_{a+1,a}{\cal D}_{--}\Si_{a+1,a}-{1\o e^2}\b{\Up}_a\Up_a \cr
+{i\o e^2}&
\b{\Phi}_{a,a+1}{\cal D}_{--}\Phi_{a,a+1}+{1\o e^2}
\bar{\La}^{\Phi}_a\La^{\Phi}_a\cr
+i~&\b{B}_{a}{\cal D}_{--}B_{a}+\bar{\La}^{B}_{a+1,a}\La^{B}_{a+1,a}\cr
+i~&\b{\til{B}}_{a}{\cal D}_{--}\til{B}_{a}
+\bar{\La}^{\til{B}}_{a+1,a}\La^{\til{B}}_{a+1,a}\cr
+\sum_{f_a=1}^{k}i~&\b{Q}_{f_aa}{\cal D}_{--}Q_{a}^{f_a}
+\bar{\La}_{f_a,a+1}^{Q}\La^{Q{f_{a}}}_{a+1}\cr
+i~&\b{\til{Q^{f_a}_a}}{\cal D}_{--}\til{Q}_{f_aa}
+\bar{\La}^{\til{Q}f_{a+1}}_{a}\La^{\til{Q}}_{f_{a+1},a} \Bigg]
}}
and the ${\cal N}=(0,2)$ superpotential term is given by
\eqn\Lagpot{\eqalign{
{\cal L}_{pot}=-\sum_{a=1}^n\int d\th^+\Bigg[~~
&\La^B_{a+1,a}(\til{B}_a\Phi_{a,a+1}-\Phi_{a,a+1}\til{B}_{a+1})\cr
+&\La^{\til{B}}_{a+1,a}(\Phi_{a,a+1}B_{a+1}-B_a\Phi_{a,a+1}) \cr
+\sum_{f_a=1}^{k}~~&
(\til{Q}_{f_aa}\Phi_{a,a+1}
-m^{f_{a+1}}_{f_a}\til{Q}_{f_{a+1},a+1})\La_{a+1}^{Q{f_a}}\cr
+&\La^{\til{Q}}_{f_{a+1},a}(\Phi_{a,a+1}Q_{a+1}^{f_{a+1}}
-Q_a^{f_a}m^{f_{a+1}}_{f_a}) +\til{Q}_{f_aa}\La^{\Phi}_aQ_a^{f_a}\cr
+&\hf t_a\Up_a|_{\bar{\th}^+=0}+\hf s_a\La^{\Phi}_a|_{\bar{\th}^+=0}
~~~\Bigg]\qquad +{c.c.}.
}}
This $(0,2)$ superpotential is easily obtained by
starting from the $(2,2)$ superpotential $W=\til{Q}(\Phi-m)Q+\tr\til{B}[\Phi,B]$,
reducing to $(0,2)$ superspace, and then orbifolding by ${\Bbb Z}_n$.
In \Lagpot, we introduced the FI-parameters and the theta-angle
\eqn\tasa{
t_a=i{\zeta^3_a\o e^2}+{\th_a\o2\pi},\quad s_a=\zeta_a^1+i\zeta_a^2.
} 

Let us explain our notation in \Lagkin\ and \Lagpot. 
We use the same letter for the $(0,2)$ chiral
superfields as the parent $(2,2)$ chiral superfields, and put the 
superscript in $\La$ for the corresponding Fermi superfields.
$(\Up_a,\La^{\Phi}_a,B_a,\til{B}_a)$ are neutral multiplets living
at the $a^{\rm th}$ node in the inner quiver (see Fig. 1). 
$(\Si_{a+1,a},\La^B_{a+1,a},\La^{\til{B}}_{a+1,a})$ are living on the link
connecting the $a^{\rm th}$ node and the $(a+1)^{\rm th}$ node 
in the inner quiver. They carry charge $(+1,-1)$ under the gauge group 
$U(1)_{a+1}\times U(1)_a$. $\Phi_{a,a+1}$ is living on the same link,
but carries opposite charge $(-1,+1)$.
$Q_a^{f_a}$ and $\til{Q}_{f_aa}$ are on the link between the $a^{\rm th}$
node in the inner quiver and the $a^{\rm th}$ node in the outer quiver;
they transform as $(+1,{\bf k})$ and $(-1,\b{\bf k})$ under 
$U(1)_a\times U(k)_a$, respectively.
Finally, the Fermi multiplet $\La^{Qf_a}_{a+1}$ is on the link 
between the $(a+1)^{\rm th}$ inner node and the $a^{\rm th}$ outer node,
and $\La^{\til{Q}}_{f_{a+1},a}$ is on the link between
the $a^{\rm th}$ inner node and the $(a+1)^{\rm th}$ outer node;
they transform as $(+1,{\bf k})$ and $(-1,\b{\bf k})$ under 
$U(1)_{a+1}\times U(k)_a$ and $U(1)_a\times U(k)_{a+1}$, respectively.

The Fermi superfields are not chiral, but satisfy the constraint
of the form $\b{{\cal D}}_+\La^i=\rt{2}E^i$.
In our case, the constraints are given by
\eqn\Laconstraint{\eqalign{
\b{{\cal D}}_+\La_a^{\Phi}&=2(\Si_{a,a-1}\Phi_{a-1,a}
-\Phi_{a,a+1}\Si_{a+1,a}-\mu_a) \cr
\b{{\cal D}}_+\La_{a+1,a}^{B}&=2(\Si_{a+1,a}B_a-B_{a+1}\Si_{a+1,a}) \cr
\b{{\cal D}}_+\La_{a+1,a}^{\til{B}}&=2(
\Si_{a+1,a}\til{B}_a-\til{B}_{a+1}\Si_{a+1,a}) \cr
\b{{\cal D}}_+\La_{a+1}^{Q{f_{a}}}&=2(\Si_{a+1,a} Q_{a}^{f_{a}}
-Q_{a+1}^{f_{a+1}}\til{m}^{f_a}_{f_{a+1}}) \cr
\b{{\cal D}}_+\La_{f_{a+1},a}^{\til{Q}}&=
-2(\til{Q}_{f_{a+1},a+1}\Si_{a+1,a}-\til{m}^{f_a}_{f_{a+1}}\til{Q}_{f_a,a}).
}}
In order for the $(0,2)$ superpotential
$-{1\o{\rt{2}}}\int d\th^+\La^iJ_i$ to be chiral, $E^i$ and $J_i$ should be
orthogonal \WittenYC
\eqn\chiralconst{
{1\o{\rt{2}}}\b{{\cal D}}_+(\La^iJ_i)=E^iJ_i=0.
}
For our choice of the superpotential \Lagpot\ and 
the constraint \Laconstraint, 
the orthogonality \chiralconst\ is satisfied by requiring
\eqn\muadef{
m^{f_a}_{f_{a-1}}\til{m}^{f_{a-1}}_{f_a'}
-\til{m}^{f_a}_{f_{a+1}}m^{f_{a+1}}_{f_a'}=\mu_a\cob^{f_a}_{f_a'}.
}
Note that $m^{f_{a+1}}_{f_a}$ and $\til{m}_{f_{a+1}}^{f_a}$ correspond to
the complex mass and the twisted mass, respectively, in the language of
${\cal N}=(2,2)$ theory. These parameters naturally live on 
the links of the outer quiver.

\subsec{Potential Energy}
We expect that the theory flows in the IR to the non-linear sigma model
with the target space given by the vanishing locus of the bosonic potential.
There are three contributions to the potential energy: the $D$-term, 
the $E$-term, and the $J$-term
\eqn\potential{
U={1\o2e^2}\sum_{a=1}^n D_a^2+\sum_i\Big(|E^i|^2+|J_i|^2\Big)
}
The $E$-term and the $J$-term can be read off from \Laconstraint\
and \Lagpot. Note that the parameter $s_a$ in \Lagpot\ enter
in the $J$-term potential as
\eqn\sapot{
2e^2\sum_{a=1}^n\lf|\til{q}_{f_aa}q^{f_a}_a+\hf s_a\ri|^2.
}
Here and hereafter we denote the scalar
component of $(0,2)$ chiral superfield by the lower case letter.
The $D$-term for the $a^{\rm th}$ $U(1)$ gauge group
is given by
\eqn\Da{
D_a=-e^2\Big(|q_a^{f_a}|^2-|\til{q}_{f_aa}|^2\Big)
+|\si_{a+1,a}|^2-|\phi_{a,a+1}|^2
-|\si_{a,a-1}|^2+|\phi_{a-1,a}|^2+\zeta^3_a.
}
In the following two sections, we will consider the moduli space of vacua
defined by $U=0$.

\newsec{Higgs Branch}
In this section, we consider the Higgs branch, {\it i.e.} the vacuum
with non-vanishing $q,\til{q}$.
For simplicity, we set $s_a=0$ and 
also set the mass term to be diagonal
\eqn\diagmass{
m_{f_{a}}^{f_{a+1}}=m_a\cob_{f_{a}}^{f_{a+1}},\quad
\til{m}_{f_{a+1}}^{f_a}=\til{m}_a\cob^{f_{a}}_{f_{a+1}}.
}
This mass term breaks the flavor symmetry $\prod_{a=1}^nU(k)_a$
down to the diagonal $U(k)$.
Then the parameter $\mu_a$ in \muadef\ is given by
\eqn\muam{
\mu_a=m_{a-1}\til{m}_{a-1}-m_a\til{m}_a.
}

Instead of setting $D_a=E^i=J_i=0$, we can consider the holomorphic condition
$E^i=J_i=0$ and mod out by the complexified gauge
group $\prod_{a=1}^n{\Bbb C}^{\times}_a$. 
Then the equation for vacuum is given by
\eqn\vaceq{\eqalign{
&\si_{a,a-1}\phi_{a-1,a}-\phi_{a,a+1}\si_{a+1,a}
=m_{a-1}\til{m}_{a-1}-m_a\til{m}_a \cr
&\phi_{a,a+1}q_{a+1}=m_aq_a,\quad\til{q}_a\phi_{a,a+1}=m_a\til{q}_{a+1} \cr
&\si_{a+1,a}q_a=\til{m}_aq_{a+1}, \quad
\til{q}_{a+1}\si_{a+1,a}=\til{m}_aq_a \cr
&\til{q}_aq_a=0.
}}
Here we suppressed the flavor index.

When all mass are non-zero, we can solve this equation up to
the gauge transformation as
\eqn\Higgsvac{\eqalign{
&\phi_{a,a+1}=m_a,\quad\si_{a+1,a}=\til{m}_a \cr
&q_a=q_{a+1},\quad\til{q}_a=\til{q}_{a+1},~~\forall a
}}
Therefore, the conditions for $(q_a,\til{q}_a)$ are reduced to
the single equation $\til{q}_1q_1=0$.
It is known that the solution space of the equation $\til{q}_1q_1=0$
modded out by ${\Bbb C}^{\times}$ gives
the $U(k)$ one-instanton moduli space on ${\Bbb R}^4$ \foot{
Turning on the FI-parameter corresponds to considering the instanton 
on non-commutative ${\Bbb R}^4$ \NekrasovSS.}.
Therefore, the Higgs branch is given by
\eqn\MHiggs{
{\cal M}_H(m,\til{m}\not=0)={\cal M}_{1,k}.
}
In the opposite extreme, {\it i.e.} when all mass are zero,
the vacuum is given by $\si_{a+1,a}=\phi_{a,a+1}=0$ and $\til{q}_aq_a=0$.
In this case, $q_a$'s and $\til{q}_a$'s with different $a$
are unrelated. Therefore, the Higgs branch is given by
\eqn\zeroHiggs{
{\cal M}_H(m=\til{m}=0)=({\cal M}_{1,k})^n.
}
We expect that in general
the dimension of the Higgs branch jumps
when setting some of the mass $m_a,\til{m}_a$ to zero.

It is natural to expect that the boundary CFT of D1-D5-KK system
is given by the Higgs branch CFT of D1-brane theory.
It is interesting to observe that the Higgs branch of our $(0,4)$
theory is basically given by the instanton moduli space, which is
the same as the Higgs branch of the D1-D5 system \WittenYU.
The similarity of the $(0,4)$ moduli space and the $(4,4)$ moduli
space is emphasized in \LarsenDH.
Our result \MHiggs, \zeroHiggs\ is consistent with the conjecture \LarsenDH\
that the D1-D5-KK system admits a solvable boundary CFT written as 
a sigma model on the symmetric product of the manifold corresponding 
to the (2345) directions.

We should also mention that there is another branch ${\cal M}_B$ 
corresponding to the expectation value of
$(B_a,\til{B}_a)$. By a similar analysis as above, we find
\eqn\Bbranch{\eqalign{
{\cal M}_B(m,\til{m}\not=0)&={\Bbb R}^4 \cr
{\cal M}_B(m=\til{m}=0)&=({\Bbb R}^4)^n.
}}
This branch represents the center-of-mass 
motion of D1-brane in the (2345) directions.

\newsec{Coulomb Branch}
In this section, we consider the Coulomb branch of our $(0,4)$ theory
and compute the 1-loop correction to the effective metric.
Let us first clarify our terminology of Coulomb and Higgs branch.
The criterion is the Higgsing of the diagonal $U(1)$ of the
gauge group $\prod_{a=1}^nU(1)_a$. In the branch with $q_a,\til{q}_a\not=0$
considered in the previous section, the diagonal $U(1)$ is broken.
On the other hand, in the branch with $\si_{a+1,a},\phi_{a,a+1}\not=0$,
the diagonal $U(1)$ is unbroken, so we will call this branch Coulomb branch.

Before going to the analysis of $(0,4)$ theory, we first review
the $(4,4)$ case.

\subsec{Case 1: D1-D5}

Let us first consider the Coulomb branch of the system of single D1-brane
and $Q_5$ D5-branes. The theory on the D1-brane is an ${\cal N}=(4,4)$
$U(1)$ gauge theory with $Q_5$ hypermultiplets with charge 1.
The Coulomb branch is parametrized by the vectormultiplet $(\Si,\Phi)$.
The metric on the Coulomb branch is corrected by integrating out the
hypermultiplet.
The 1-loop corrected metric is
given by \refs{\DouglasVU,\DiaconescuGU}
\eqn\cmet{
ds^2=\lf({1\o e^2}+{Q_5\o |\phi|^2+|\si|^2}\ri)(|d\phi|^2+|d\si|^2).
}
From the general structure of ${\cal N}=(4,4)$ 
non-linear sigma model, it can be shown that the metric is 1-loop exact.
One can immediately notice that \cmet\ is the metric on the transverse space
of $Q_5$ NS5-branes \CallanAT. 
Near the origin of Coulomb branch, this metric reduces
to the familiar throat metric, and the CFT on the Coulomb branch
is described by 
\eqn\throat{
{\cal M}_C^{\rm near~horizon}={\Bbb R}_\phi\times SU(2)_{Q_5}
}
where ${\Bbb R}_\phi$ is the linear dilaton CFT and $SU(2)_{Q_5}$
is the WZW model.

\subsec{Case 2: D1 on ${\Bbb C}^2/{\Bbb Z}_n$}

The gauge theory on the D1-brane on ${\Bbb C}^2/{\Bbb Z}_n$
is given by the ${\cal N}=(4,4)$ $\h{A}_{n-1}$ quiver gauge theory
\refs{\DouglasSW,\JohnsonPY}.
In our notation, this is given by the matter content 
on the inner quiver in Fig. 1.
The fields written in the $(0,2)$ superspace naturally organize themselves
into the ${\cal N}=(4,4)$ multiplets. Namely, 
$(B_a,\til{B}_a,\Up_a,\La^{\Phi}_a)$ is the $(4,4)$ vectormultiplet living
on the $a^{\rm th}$ node, and
$(\Si_{a+1,a},\Phi_{a,a+1},\La^B_{a+1,a},\La^{\til{B}}_{a+1,a})$ is the 
$(4,4)$
hypermultiplet living on the link between the $a^{\rm th}$ node and 
the $(a+1)^{\rm th}$ node. 
In this case, the diagonal $U(1)$ is decoupled and it never gets Higgsed.
Although it is natural to call the branch of non-zero 
$(\si_{a+1,a},\phi_{a,a+1})$ Coulomb branch in the presence of D5-brane,
this branch is usually referred to as the 
Higgs branch in the case of D1 on ${\Bbb C}^2/{\Bbb Z}_n$,
since it is parametrized by the hypermultiplet.

It is known that the metric on the Higgs branch is the ALE metric. 
This is obtained by the following steps \GibbonsNT. We
first note that the action of the $(4,4)$ quiver theory is 
reproduced from our $(0,4)$ quiver theory
by setting the fields from the 1-5 string to zero.
In particular, the equations for the vacuum 
\eqn\vaceq{\eqalign{
\si_{a,a-1}\phi_{a-1,a}-\si_{a+1,a}\phi_{a,a+1}&=\mu_a \cr
|\si_{a,a-1}|^2-|\phi_{a-1,a}|^2
-|\si_{a+1,a}|^2+|\phi_{a,a+1}|^2&=\zeta_a^3.
}}  
is the same as those for the Coulomb branch of our $(0,4)$ theory.
For the consistency, the FI-parameters have to satisfy
the condition
\eqn\sumzero{
\sum_{a=1}^n\mu_a=\sum_{a=1}^n\zeta_a^3=0.
}
Next we introduce the variables $(\vec{r}_a,\varphi_a)$ by
\eqn\mommap{\eqalign{
2\si_{a+1,a}\phi_{a,a+1}&=r_a^1+ir_a^2 \cr
|\si_{a+1,a}|^2-|\phi_{a,a+1}|^2&=r_a^3 \cr
\arg(\si_{a+1,a})&=\varphi_a.
}}
The vacuum equations \vaceq\ imply that
\eqn\rainr{
\vec{r}_a=\vec{r}-\vec{x}_a
}
where $\vec{x}_a$ is given by
\eqn\xvec{
\vec{x}_a=\sum_{b=1}^a\vec{\mu}_b,\qquad 
\vec{\mu}_a=(2{\rm Re}\mu_a,2{\rm Im}\mu_a,\zeta_a^3).
}
Then we write down the kinetic term of $(\si_{a+1,a},\phi_{a,a+1})$
restricted on the locus \vaceq
\eqn\kinhyp{\eqalign{
{\cal L}_{kin}&=-{1\o e^2}\sum_{a=1}^n\Big[|D_\mu\si_{a+1,a}|^2
+|D_\mu\phi_{a,a+1}|^2\Big] \cr
&=-{1\o e^2}\sum_{a=1}^n\lf[{1\o |\vec{r}-\vec{x}_a|}(\del_\mu\vec{r})^2
+|\vec{r}-\vec{x}_a|(\del_\mu\varphi_a+\vec{\om}_a\cdot\del_\mu\vec{r}
+A_{a,\mu}-A_{a+1,\mu})^2\ri]
}}
where $\vec{\om}_a$ satisfies
\eqn\diracom{
\nabla\times\vec{\om}_a=\hf\nabla\lf({1\o|\vec{r}-\vec{x}_a|}\ri)
}
Finally, by classically integrating out the gauge field $A_{a,\mu}$,
we arrive at the effective Lagrangian on the locus \vaceq
\eqn\ALEkin{
{\cal L}_{eff}=-{1\o e^2}\Big[V(\del_\mu\vec{r})^2+V^{-1}
(\del_\mu\th+\vec{\om}\cdot\del_\mu\vec{r})^2\Big]
}
where $V,\vec{\om}$ and $\th$ are given by
\eqn\Vomth{
V=\sum_{a=1}^n{1\o |\vec{r}-\vec{x}_a|},\quad
\vec{\om}=\sum_{a=1}^n\vec{\om}_a,\quad
\th=\sum_{a=1}^n\varphi_a.
}
The metric \ALEkin\ is nothing but the familiar ALE metric.
The above procedure is known as the hyperK\"{a}hler quotient
construction of ALE metric.

\subsec{Supergravity Analysis}
Now let us go back to the analysis of $(0,4)$ theory.
From the discussion in the case 1,2 above, we can expect that the 
metric on the Coulomb branch of our $(0,4)$ theory reproduces
the supergravity solution of D1-D5-KK system, which is given by the usual
harmonic function rule \JohnsonGA
\eqn\DKKmet{\eqalign{ 
ds^2&={1\o\rt{Z_1Z_5}}(-dt^2+dx_1^2)+\rt{Z_1\o Z_5}(dx_2^2+\cdots dx_5^2)
+\rt{Z_1Z_5}ds^2_{KK}\cr
e^{-2\Phi}&={Z_5\o Z_1},\qquad Z_{1,5}=1+{Q_{1,5}\o r}.
}}
Here $ds^2_{KK}$ denotes the metric of KK-monopoles, 
{\it i.e.} Taub-NUT space. The metric seen by the probe D1-brane 
can be obtained by plugging the solution \DKKmet\ into the D1-brane 
Born-Infeld action $S_{D1}=-\int d\tau d\si e^{-\Phi}\rt{-\det g}$, 
and expanding it around the configuration 
$X^0=\tau, X^1=\si, X^m=X^m(\tau)$, where
$X^m$ is the coordinate on the Taub-NUT space.
By looking at the term quadratic in the velocity $\dot{X}^m(\tau)$, we find
that the effective metric seen by the D1-brane probe is
\eqn\dsprobe{
ds^2_{{\rm D1-probe}}=Z_5ds^2_{KK}.
}
This is independent of the factor $Z_1$ as expected from the BPS property.

The metric \dsprobe\ can also be obtained as the string metric
on the S-dual F1-NS5-KK system. The S-dual of \DKKmet\ is given by
\eqn\NSKKmet{\eqalign{
&ds^{2}{'}=
e^{-\Phi}ds^2=
 {1\o Z_1}(-dt^2+dx_1^2)+dx_2^2+\cdots dx_5^2
+Z_5ds^2_{KK} \cr
&e^{-2\Phi'}=e^{2\Phi}={Z_1\o Z_5}.
}}
Therefore, the metric on the transverse space seen by the fundamental
string is
\eqn\NSfKK{
ds^2_{\rm NS5-KK}=Z_5ds^2_{KK},
}
which agrees with the D1-brane probe metric \dsprobe\ as expected.
In the next subsection, we compute the 1-loop correction to
the metric on the Coulomb branch and compare it with the 
supergravity result \NSfKK.

\subsec{One-loop Correction to the Quiver Gauge Theory}
In this section, we compute the 1-loop correction to the metric
on the Coulomb branch. 
There is no correction from $(B,\til{B})$, 
since if we turn off the fields  
coming from the 1-5 string, the theory reduces to the ${\cal N}=(4,4)$
quiver gauge theory and it is known that there is no correction
to the Higgs branch metric in this case. Recall that, as we discussed
in section 4.2, the role of Coulomb and Higgs branch is exchanged
in the D1-brane on ${\Bbb C}^2/{\Bbb Z}_n$ case.
Therefore, the 1-loop correction only comes from integrating out the modes of
1-5 string, {\it i.e.} the fields $(Q_a^{f_a},\til{Q}_{f_aa},\La^{Qf_a}_{a+1},
\La^{\til{Q}}_{f_{a+1},a})$.
In this section,
we set $m=\til{m}=s_a=0$ for simplicity.
Then the equation for vacuum is given by \vaceq\ with $\mu_a=0$.

Let us compute the 1-loop integral in the background
of $(\si_{a+1,a},\phi_{a,a+1})$ obeying \vaceq. 
The term quadratic in bosonic fields is given by
\eqn\Lbose{\eqalign{
{\cal L}_{\rm boson}=-\sum_{a=1}^n\Big(~&
|D_\mu q_a|^2+2(|\si_{a+1,a}|^2+|\phi_{a-1,a}|^2)|q_a|^2 \cr
+&|D_\mu\til{q_a}|^2+2(|\si_{a-1,a}|^2+|\phi_{a,a+1}|^2)|\til{q}_a|^2~\Big)
}}
and the fermionic term is given by
\eqn\fermikin{
{\cal L}_{\rm fermi}=\sum_{a=1}^n\Psi_a^\dag\lf(\matrix{i\del_{--}&M_a
\cr M_a^\dag&i\del_{++}}\ri)\Psi_a
}
where $\Psi_a$ and $M_a$ are defined as
\eqn\defMpsi{
\Psi_a=\lf(\matrix{\psi_{+,a}\cr\b{\til{\psi}}_{+,a}\cr\psi_{-,a+1}\cr
\b{\til{\psi}}_{-,a-1}}\ri),\qquad
M_a=\rt{2}\lf(\matrix{-\b{\si}_{a+1,a}&\b{\phi}_{a-1,a}\cr
\phi_{a,a+1}&\si_{a,a-1}}\ri).
}
Here we suppressed the flavor indices for notational simplicity.
We can check that the bosons and fermions have the same mass eigenvalues
\eqn\MMdag{\eqalign{
M_aM_a^\dag&=2\lf(\matrix{|\si_{a+1,a}|^2+|\phi_{a-1,a}|^2&
-\b{\si}_{a+1,a}\b{\phi}_{a,a+1}+\b{\si}_{a,a-1}\b{\phi}_{a-1,a}\cr
-\si_{a+1,a}\phi_{a,a+1}+\si_{a,a-1}\phi_{a-1,a}
&|\si_{a-1,a}|^2+|\phi_{a,a+1}|^2
}\ri) \cr
&=2\lf(\matrix{|\si_{a+1,a}|^2+|\phi_{a-1,a}|^2&0\cr
0&|\si_{a-1,a}|^2+|\phi_{a,a+1}|^2}\ri).
}}
In the last step, we used the vacuum equation \vaceq\
with $\mu_a=0$. Clearly, the eigenvalues of $M_aM_a^\dag$
agree with the mass of $q_a,\til{q}_a$.

Now we can compute the 1-loop effective action
by integrating out the 1-5 string modes:
\eqn\Galoop{\eqalign{
\Ga_{\rm 1-loop}=k\sum_{a=1}^n\Bigg\{&\log\det\Big[\del_{--}\del_{++}
+2(|\si_{a+1,a}|^2+|\phi_{a-1,a}|^2)\Big] \cr
+&\log\det\Big[\del_{--}\del_{++}
+2(|\si_{a-1,a}|^2+|\phi_{a,a+1}|^2)\Big] \cr
-&\log\det\lf(\matrix{i\del_{--}&M_a
\cr M_a^\dag&i\del_{++}}\ri) ~\Bigg\} \cr
=-k\sum_{a=1}^n\Tr&\lf[
{1\o (\del_{--}\del_{++}+M_aM^\dag_a)^2}
\del_{++}M_a\del_{--}M_a^\dag
\ri]+\cdots.
}}
Here we expanded the result up to two derivatives and the dots denote
the higher derivative terms. The 1-loop integral in 
\Galoop\ can be evaluated by using the formula
\eqn\intp{
\int{d^2p\o(2\pi)^2}{1\o (p^2+m^2)^2}={1\o4\pi m^2}.
}
Finally, we arrive at the 1-loop corrected 
effective Lagrangian
\eqn\efflag{\eqalign{
{\cal L}_{\rm eff}=-\sum_{a=1}^n\Bigg[&{1\o e^2}
\Big(|D_\mu\si_{a+1,a}|^2+|D_\mu\phi_{a,a+1}|^2\Big)
\cr
+&{k\o4\pi}\lf({|D_\mu\si_{a+1,a}|^2+|D_\mu\phi_{a-1,a}|^2
\o|\si_{a+1,a}|^2+|\phi_{a-1,a}|^2}+
{|D_\mu\si_{a,a-1}|^2+|D_\mu\phi_{a,a+1}|^2
\o|\si_{a,a-1}|^2+|\phi_{a,a+1}|^2}\ri)\Bigg].
}}
The first line in \efflag\ is the tree level term and the second line
is the 1-loop correction.

For the general value of the FI parameters $\zeta_a^3$,
it is not so easy to rewrite this Lagrangian in terms of the
variables $(\vec{r},\th)$ introduced in section 4.2.
Here we focus on the orbifold limit corresponding to $\mu_a=\zeta_a^3=0$.
Then we can show that
\eqn\orbrel{
|\si_{a+1,a}|^2+|\phi_{a-1,a}|^2=|\si_{a,a-1}|^2+|\phi_{a,a+1}|^2
=|\vec{r}|\equiv r\quad\forall a.
}
This implies that the denominators in the second line in \efflag\ are 
common for all terms and they can be factored out.
Therefore, the 1-loop correction term in \efflag\ becomes proportional
to the tree level term. In other words, the metric on the Coulomb branch is
given by the ALE metric up to a conformal factor: 
\eqn\dsloop{
ds^2_{\rm 1-loop}=\lf({1\o e^2}+{k\o2\pi r}\ri)
\Big({n\o r}(d\vec{r})^2+{r\o n}
(d\th+\vec{\om}\cdot d\vec{r})^2\Big).
}
This is exactly the metric obtained in the supergravity approximation
\NSfKK, with the understanding that the KK-monopole metric $ds^2_{KK}$
is replaced by the orbifold limit of the ALE metric.
This shows that our $(0,4)$ theory can be thought of as a GLSM
for the NS5-branes on ALE space.
Near the origin of Coulomb branch, the metric \dsloop\ reduces to
\eqn\nearHmet{
ds^2_{\rm 1-loop}\sim{kn\o2\pi}\lf[{dr^2\o r^2}+d\Om^2_{\S^2}
+{1\o n^2}(d\th+\vec{\om}\cdot d\vec{r})^2\ri].
}
One can easily show that the last two terms is the metric on the lens
space $\S^3/{\Bbb Z}_n$ written as the Hopf fibration of $\S^1$
over $\S^2$.
Therefore, the near horizon limit is described by
the ${\Bbb Z}_n$ orbifold of throat CFT
\eqn\nearHcft{
{\cal M}_C^{\rm near~horizon}={\Bbb R}_\phi\times {SU(2)_{kn}\o{\Bbb Z}_n}
}
where $\phi=\log r$ is the linear dilaton direction.

\newsec{Discussion}
In this paper, we have constructed
the ${\cal N}=(0,4)$ quiver gauge theory corresponding to the D1-D5 branes on 
${\Bbb C}^2/{\Bbb Z}_n$. The 1-loop correction of the Coulomb branch
shows that this theory can be seen as a GLSM of NS5-branes on ALE space.
Our result of 1-loop effective action on the Coulomb branch
\efflag\ is not proportional
to the tree level term for the general FI parameters.
This seems to suggest that the naive harmonic function rule breaks down in
the string theory. We also expect that the effective metric for the $(0,4)$
theory is not 1-loop exact, although it is constrained by supersymmetry to be
the hyperK\"{a}hler with torsion sigma model.
For both Higgs and Coulomb branch, our analysis is limited to the special
value of the FI parameters. It would be interesting to study the
general parameter case.

\vskip10mm
\noindent{\bf Acknowledgment:}
I would like to thank Yuji Sugawara for discussion in 1998.
\listrefs
\bye